\documentclass[12pt]{article}

\usepackage[margin=1in]{geometry}

\usepackage{setspace}

\usepackage{graphicx}

\usepackage{enumitem}

\usepackage{microtype}
\usepackage{subfigure}
\usepackage{booktabs}
\usepackage{comment}
\usepackage{wrapfig}
\usepackage{arydshln}
\usepackage{enumitem}
\usepackage{multirow}
\usepackage{url}
\usepackage{natbib}
\usepackage{authblk}

\RequirePackage[colorlinks,citecolor=blue,linkcolor=blue,urlcolor=blue,pagebackref]{hyperref}

\usepackage{amsmath}
\usepackage{amssymb}
\usepackage{mathtools}
\usepackage{amsfonts}
\usepackage{bm}
\usepackage{bbm}

\usepackage{algorithm}
\usepackage{algorithmic}

\def\eqref#1{equation~\ref{#1}}

\def\1{\bm{1}}

\DeclareMathAlphabet{\mathsfit}{\encodingdefault}{\sfdefault}{m}{sl}
\SetMathAlphabet{\mathsfit}{bold}{\encodingdefault}{\sfdefault}{bx}{n}

\def\calG{{\mathcal{G}}}

\def\calI{{\mathcal{I}}}

\def\calS{{\mathcal{S}}}

\def\calX{{\mathcal{X}}}

\def\bbE{{\mathbb{E}}}

\def\bbR{{\mathbb{R}}}

\DeclareMathOperator*{\argmax}{arg\,max}
\DeclareMathOperator*{\argmin}{arg\,min}

\newcommand{\sqb}[1]{\left[#1\right]}
\newcommand{\cb}[1]{\left\{#1\right\}}

\newcommand{\bigp}[1]{\big(#1\big)}

\newcommand{\Bigp}[1]{\Big(#1\Big)}
\newcommand{\Bigsqb}[1]{\Big[#1\Big]}

\usepackage{amsthm}

\theoremstyle{plain}

\newtheorem{theorem}{Theorem}[section]

\newtheorem*{remark}{Remark}

\usepackage[textsize=tiny]{todonotes}
\usepackage{multirow}
\usepackage{wrapfig}
\usepackage{subfigure}

\renewcommand{\eqref}[1]{(\ref{#1})}

\usepackage{xcolor}
\newcount\Comments  
\newcommand{\kibitz}[2]{\ifnum\Comments=1\textcolor{#1}{#2}\fi}

\usepackage[capitalize,noabbrev]{cleveref}

\allowdisplaybreaks

\title{Causal-Policy Forest for End-to-End Policy Learning}

\author{Masahiro Kato\thanks{Email: \texttt{mkato-csecon@g.ecc.u-tokyo.ac.jp}}$\,$}

\affil{The University of Tokyo}

\date{\today}

\begin{document}

\maketitle 

\begin{abstract}
    This study proposes an end-to-end algorithm for policy learning in causal inference. We observe data consisting of covariates, treatment assignments, and outcomes, where only the outcome corresponding to the assigned treatment is observed. The goal of policy learning is to train a policy from the observed data, where a policy is a function that recommends an optimal treatment for each individual, to maximize the policy value. In this study, we first show that maximizing the policy value is equivalent to minimizing the mean squared error for the conditional average treatment effect (CATE) under $\{-1, 1\}$ restricted regression models. Based on this finding, we modify the causal forest, an end-to-end CATE estimation algorithm, for policy learning. We refer to our algorithm as the causal-policy forest. Our algorithm has three advantages. First, it is a simple modification of an existing, widely used CATE estimation method, therefore, it helps bridge the gap between policy learning and CATE estimation in practice. Second, while existing studies typically estimate nuisance parameters for policy learning as a separate task, our algorithm trains the policy in a more end-to-end manner. Third, as in standard decision trees and random forests, we train the models efficiently, avoiding computational intractability.
\end{abstract}

\section{Introduction}
The decision-making problem in causal inference is often defined as a social welfare maximization problem, where the goal is to recommend the best treatment for each individual. This problem is called policy learning, and we pursue this goal by training a policy, a function that takes covariates and returns a treatment recommendation, using observational data \citep{Swaminathan2015counterfactualrisk,Swaminathan2015batchlearning,Kitagawa2018whoshould,Athey2021policylearning,Zhou2023offlinemultiaction}. This study proposes an end-to-end policy learning algorithm using random forests, which we call the causal-policy forest. While existing methods often require estimating an (asymptotically) unbiased target welfare and then training a policy by maximizing the estimated welfare, our method integrates these steps in a unified way. Our algorithm also bridges conditional average treatment effect (CATE) estimation and policy learning. Although they have been treated as different tasks, we show that they are closely connected, and that CATE estimation methods can often be adapted for policy learning.

Given covariates, the optimal policy recommends the treatment with the highest conditional expected outcome. However, the conditional expected outcome is usually unknown. To address this challenge, two main approaches have been proposed, empirical welfare maximization (EWM) and plug-in policy approaches \citep{Kitagawa2018whoshould}. EWM is also referred to as counterfactual risk minimization. In the EWM approach, we first approximate the target welfare and then train a policy to maximize social welfare. In contrast, the plug-in policy approach first estimates the conditional expected outcome and then recommends the treatment with the highest estimated conditional expected outcome. When treatments are binary, the first stage of the plug-in approach can be interpreted as CATE estimation, which is another core problem in causal inference.

This study first shows that the CATE estimation objective can be interpreted as an EWM objective. We show that if we regress the treatment effect using regression models restricted to $\{-1, 1\}$ values, then the resulting regression function estimator corresponds to an optimal policy. Based on this finding, we design our policy learning algorithm as a modification of the causal forest, which \citet{Wager2018estimationinference} designs for CATE estimation. We refer to our algorithm as the causal-policy forest. The proposed algorithm can train a policy directly without explicitly estimating welfare, unlike existing policy learning methods such as \citet{Kitagawa2018whoshould}, \citet{Athey2021policylearning}, and \citet{Zhou2023offlinemultiaction}.

\section{Setup}
This section describes our problem setting. For simplicity, we focus on the case with binary treatments, but it is easy to extend our results to the multiple treatment case.

\paragraph{Observations.}
Let the sample size be $n$. We observe i.i.d.\ draws $Z_i=(Y_i,D_i,X_i)$, $i=1,\dots,n$, where $X_i\in\calX\subset\bbR^{d_\calX}$ is a vector of pre-treatment covariates, $D_i\in\{0,1\}$ is a binary treatment indicator, and $Y_i\in\bbR$ is the observed outcome.

\paragraph{Potential outcomes.}
Under the Neyman Rubin model, each unit has potential outcomes $(Y_{0,i},Y_{1,i})$ and $Y_i=D_i Y_{1,i}+(1-D_i)Y_{0,i}$. We assume \emph{unconfoundedness}, $(Y_{0,i},Y_{1,i})\perp\perp D_i \mid X_i$, and define $P$ as the population distribution of $(Y_{0},Y_{1},D,X)$.

\paragraph{Policy and welfare.}
Let us denote a class of policies $\pi:\calX\to[0,1]$ by $\Pi$. We then define the (utilitarian) social welfare of $\pi \in \Pi$ as
\[
W(\pi)\coloneqq \bbE\bigl[Y_1\pi(X)+Y_0(1-\pi(X))\bigr].
\]
Our goal is to learn the optimal policy from the observations. It maximizes social welfare and is denoted as
\[\pi^*_\Pi \coloneqq \argmax_{\pi \in \Pi} W(\pi).\]
When the dependency is clear from the context, we omit $\Pi$ and simply denote it as $\pi^*$. 
We denote our trained policy as $\widehat{\pi}$. 

\paragraph{Regret.}
We assess $\widehat\pi$ via the regret, defined as
\[
\mathrm{Regret}\bigl(\pi^*_\Pi,\widehat\pi\bigr)\coloneqq W(\pi^*_\Pi)-W(\widehat\pi),
\qquad
\pi^*_\Pi \in \argmax_{\pi\in\Pi} W(\pi).
\]

\paragraph{First-best policy.}
If $\Pi$ is the set of all measurable functions $\pi:\calX\to[0,1]$, the optimal policy is given as
\[
\pi^*_{\mathrm{FB}}(x)=\mathbbm{1}[\tau_0(x)\ge 0] = \argmax_{\pi\in\Pi} W(\pi),
\]
where $\tau_0(x)\coloneqq \bbE[Y_1-Y_0\mid X=x]$ denotes the CATE.
Following \citet{Kitagawa2018whoshould}, we call such a policy the first-best policy. For general $\Pi$, which might not include $\pi^*_{\mathrm{FB}}$, we denote the best policy in $\Pi$ as $\pi^*_\Pi$.

\subsection*{Notation and assumptions}
Let $e(x)=P(D=1\mid X=x)$ be the propensity score and $m_d(x)=\bbE[Y_d\mid X=x]$ the conditional mean outcome. We assume that there exists a constant $0< \epsilon< 1/2$ independent of $n$ such that $\epsilon < e(X) < 1 - \epsilon$ almost surely. We also assume that the variables $X$ and $Y$ are bounded.

\section{CATE Reduction of Policy Learning}
\label{sec:equivalence}
We now formalize the core equivalence. Let $\Pi$ be a set of deterministic policies; that is, $\pi(x) \in \{1, 0\}$.\footnote{We can generalize this result for a more general policy. Since it is not our main interest, we omit it. For the details, see \citet{Kato2025bridginggap}.} Throughout this section, let $\calG_\Pi\coloneqq\{g=2\pi-1:\pi\in\Pi\}$. We first consider the case where $\pi$ is binary valued; that is, $\pi \colon \calX \to \{0, 1\}$, where $\pi(X) = 1$ (resp.\ $\pi(X) = 0)$ means that the policy recommends treatment $1$ (resp.\ treatment $0$) for given covariates $X$.

We show the equivalence between the EWM and plug-in approaches in an idealized setting where we can observe $Y_1$ and $Y_0$ directly, so we can ignore the counterfactual nature.
Recall that the welfare maximization problem can be written as
\begin{align*}
    \max_{\pi\in\Pi} \bbE\sqb{\pi(X)Y_1 + (1 - \pi(X))Y_0}.
\end{align*}

In this case, we consider the following regression problem:
\begin{align*}
    &\min_{g \in \calG_\Pi} \bbE\sqb{\Bigp{\tau_0(X) - g(X)}^2},
\end{align*}
where $\calG_\Pi \coloneqq \{g = (2\pi - 1) \colon \pi \in \Pi\}$. By definition, $g$ is a function such that $g\colon \calX \to \{-1, 1\}$.

We now show the equivalence between EWM and least squares.
Recall that we defined the optimal policy $\pi^*$ as
\begin{align*}
    \pi^* = \argmax_{\pi\in\Pi} W(\pi) = \argmax_{\pi\in\Pi} \bbE\sqb{\pi(X)Y_1 + (1 - \pi(X))Y_0}.
\end{align*}

Let $g^*$ be the optimal predictor defined as
\begin{align*}
    g^* \coloneqq \argmin_{g \in \calG_\Pi}\bbE\sqb{\Bigp{\tau_0(X) - g(X)}^2}.
\end{align*}

Then, the following theorem holds.
\begin{theorem}
\label{thm:popequivalence}
It holds that
\[g^* = 2\pi^* - 1.\]
\end{theorem}
The proof is straightforward and is given below.

\begin{proof}
For $\calG_\Pi$, we have
\begin{align*}
    g^* &= \argmin_{g \in \calG_\Pi}\bbE\sqb{\Bigp{\tau_0(X) - g(X)}^2}\\
    &= \argmin_{g \in \calG_\Pi}\bbE\sqb{\Bigp{\bigp{Y_1 - Y_0} - g(X)}^2}\\
    &= \argmin_{g \in \calG_\Pi}\bbE\sqb{\bigp{Y_1 - Y_0}^2 - 2g(X)\bigp{Y_1 - Y_0} + g(X)^2}\\
    &= \argmin_{g \in \calG_\Pi}\bbE\sqb{ - 2g(X)\bigp{Y_1 - Y_0} + g(X)^2}\\
    &= \argmin_{g \in \calG_\Pi}\bbE\sqb{ - 2g(X)\bigp{Y_1 - Y_0} + 1}\\
    &= \argmin_{g \in \calG_\Pi}\bbE\sqb{ - 2g(X)\bigp{Y_1 - Y_0} }.
\end{align*}
Here, we used $g(X)^2 = 1$. We omitted $\bigp{Y_1 - Y_0}^2$ from the third line to the forth line and $1$ from the fifth line to the sixth line, since they are irrelevant to the optimization.

Continuing, we have
\begin{align*}
    g^* = &\argmin_{g \in \calG_\Pi}\bbE\sqb{ - 2g(X)\bigp{Y_1 - Y_0} }\\
    &= \argmin_{g \in \calG_\Pi}\bbE\sqb{ - 2\bigp{g(X) + 1}\bigp{Y_1 - Y_0} + 2 \bigp{Y_1 - Y_0}}.
\end{align*}
We added and subtracted terms that are irrelevant to the optimization. Here, recall that
\begin{align*}
    \pi^* &= \argmin_{\pi \in \cb{\pi(\cdot) = \bigp{g(\cdot) + 1}/2 \colon g \in \calG_\Pi}}\bbE\sqb{ - \pi(X)Y_1 - (1 - \pi(X))Y_0}\\
    &= \argmin_{\pi \in \cb{\pi(\cdot) = \bigp{g(\cdot) + 1}/2 \colon g \in \calG_\Pi}}\bbE\sqb{ - \pi(X)(Y_1 - Y_0) - Y_0}.
\end{align*}
Therefore, $g^* = 2\pi^* - 1$ holds, and the proof is complete.
\end{proof}

This theorem implies that the EWM approach is equivalent to least squares, where we regress $Y_1 - Y_0$ using a function $g \colon \calX \to \{-1, 1\}$.
\section{Causal-Policy Forest}
By utilizing the equivalence between CATE estimation and policy learning, we propose an end-to-end policy learning algorithm, called causal-policy forest.

\subsection{Basic Idea}
Our proposed algorithm is a modification of causal forest, originally proposed in \citet{Wager2018estimationinference} for CATE estimation.
Causal forest ensembles causal trees that split the covariate space by recursive partitioning \citep{Athey2016recursivepartitioning}, and estimate the CATE within each leaf.
In its simplest form, when $x \in \calX$ belongs to a leaf $\calS \subset \calX$, a causal tree returns the leafwise difference in mean outcomes,
\begin{align*}
    \widehat{\tau}(x)
    &\coloneqq
    \frac{1}{\sum^n_{i=1}\mathbbm{1}[X_i \in \calS, D_i = 1]}
    \sum^n_{i=1}\mathbbm{1}[X_i \in \calS]D_iY_i
    -
    \frac{1}{\sum^n_{i=1}\mathbbm{1}[X_i \in \calS, D_i = 0]}
    \sum^n_{i=1}\mathbbm{1}[X_i \in \calS](1 - D_i)Y_i,
\end{align*}
and causal forest averages these estimates over many trees.
Causal trees place splits so that the resulting piecewise constant estimator attains small squared error for the CATE target.

Causal-policy forest modifies this construction for policy learning.
Recall that for binary treatments, a deterministic policy can be represented by $g(x)=2\pi(x)-1 \in \{-1,1\}$, where $g(x)=1$ corresponds to recommending treatment $1$ and $g(x)=-1$ corresponds to recommending treatment $0$.
Our causal-policy tree outputs a leafwise policy score by taking the sign of the leafwise CATE estimate,
\begin{align}
\label{predict_cate}
    \widehat{g}(x)
    \coloneqq
    \begin{cases}
        1  & \text{if }\widehat{\tau}(x) \ge 0,\\
        -1 & \text{if }\widehat{\tau}(x) < 0.
    \end{cases}
\end{align}
Both in the splitting rule and final outputs, instead of targeting the squared error of $\widehat{\tau}(x)$ as a real valued predictor, we target the squared error of the restricted predictor $\widehat{g}(x)\in\{-1,1\}$ for the CATE target.

This choice of $\widehat{g}(x)$ corresponds to solving the following MSE minimization problem:
\begin{align*}
    &\min_{g \in \calG_\Pi} \bbE\sqb{\Bigp{\tau_0(X) - g(X)}^2}.
\end{align*}
This is because $\widehat{\tau}(x)$ is an estimator of $\tau_0(x)$, and choosing $\widehat{g}(x)$ as in \eqref{predict_cate} minimizes the squared error between $\widehat{\tau}(x)$ and $\widehat{g}(x)$ in the sense that
\begin{align*}
    \widehat{g}(x) = \argmin_{g(x) \in \{-1, 1\}} \Bigp{\widehat{\tau}(x) - g(x)}^2.
\end{align*}

\begin{remark}[Tree-based Riesz regression]
\label{rem:treeriesz}
Note that the CATE estimator $\widehat{\tau}(x)$ can be written as
\begin{align*}
    \widehat{\tau}(x)
    &=
    \frac{1}{n}
    \sum^n_{i=1}\widehat{\alpha}(X_i)Y_i,
\end{align*}
where
\[\widehat{\alpha}(X_i) \coloneqq D_i\frac{\mathbbm{1}[X_i \in \calS]}{\frac{1}{n}\sum^n_{i=1}\mathbbm{1}[X_i \in \calS, D_i = 1]}
    - (1 - D_i)\frac{\mathbbm{1}[X_i \in \calS]}{\frac{1}{n}\sum^n_{i=1}\mathbbm{1}[X_i \in \calS, D_i = 0]}.\]
Here, $\widehat{\alpha}(X_i)$ corresponds to an estimator of the Riesz representer \citep{Chernozhukov2021automaticdebiased,Chernozhukov2022automaticdebiased}.
This Riesz representer estimation is a variant of Riesz regression.
This type of estimator construction in Riesz regression shares the same idea as nearest neighbor matching, as pointed out in \citet{Lin2023estimationbased} and \citet{Kato2025nearestneighbor}.
Unlike nearest neighbor matching, we construct the kernel function through recursive partitioning in Riesz regression.
This observation suggests that causal trees and causal forests endogenize the Riesz regression estimation procedure from the CATE estimation perspective.
\end{remark}

\begin{remark}[Empirical welfare]
    Remark~\ref{rem:treeriesz} and results in \citet{Lin2023estimationbased} and \citet{Kato2025nearestneighbor} imply that
    $\frac{1}{\widehat e(\cdot)} \approx \frac{\mathbbm{1}[X_i \in \calS]}{\frac{1}{n}\sum^n_{i=1}\mathbbm{1}[X_i \in \calS, D_i = 1]}$ and $\frac{1}{1 - \widehat e(\cdot)} \approx \frac{\mathbbm{1}[X_i \in \calS]}{\frac{1}{n}\sum^n_{i=1}\mathbbm{1}[X_i \in \calS, D_i = 0]}$ hold. For details, see \citet{Lin2023estimationbased} and \citet{Kato2025nearestneighbor}. Thus, $\widehat{\tau}(x)$ corresponds to an inverse probability weighting estimator of the CATE, which is also used as empirical welfare in \citet{Kitagawa2018whoshould}. 
\end{remark}

\subsection{Algorithm}
We now describe the causal-policy forest in detail.
The construction follows the honest forest design, where the data used to place splits are separated from the data used to estimate leaf quantities.
This separation is important in our setting because the final output is binary valued, and adaptive reuse of the same observations for splitting and estimation can lead to unstable sign decisions.
Throughout, we place splits by directly minimizing an MSE criterion that matches our prediction space, namely $g(x)\in\{-1,1\}$.

\paragraph{Inputs.}
Let $B$ be the number of trees, $s$ be the subsample size per tree, and $k$ be the minimum leaf size per treatment arm.
Let $m$ be the number of candidate split variables considered at each node.

\paragraph{Tree construction.}
For each $b=1,\dots,B$, repeat the following steps.
\begin{enumerate}
\item Draw a subsample $\calI_b \subset \{1,\dots,n\}$ of size $s$ without replacement, and split it into two disjoint sets,
$\calI_b^{\mathrm{est}}$ and $\calI_b^{\mathrm{split}}$, with $|\calI_b^{\mathrm{est}}| \approx |\calI_b^{\mathrm{split}}| \approx s/2$.

\item Grow a binary tree by recursive partitioning using the split sample $\calI_b^{\mathrm{split}}$.
At each node corresponding to a region $\calS \subset \calX$, we draw $m$ candidate split variables at random.
For each candidate split (variable and split point) that partitions $\calS$ into two children $\calS_L$ and $\calS_R$, define the split-sample child-wise CATE estimates
\begin{align*}
    \widehat{\tau}^{\mathrm{split}}_{b}(\calS')
    \coloneqq
    \frac{\sum_{i\in\calI_b^{\mathrm{split}}}\mathbbm{1}[X_i\in\calS', D_i=1]Y_i}{\sum_{i\in\calI_b^{\mathrm{split}}}\mathbbm{1}[X_i\in\calS', D_i=1]}
    -
    \frac{\sum_{i\in\calI_b^{\mathrm{split}}}\mathbbm{1}[X_i\in\calS', D_i=0]Y_i}{\sum_{i\in\calI_b^{\mathrm{split}}}\mathbbm{1}[X_i\in\calS', D_i=0]},
    \qquad \calS'\in\{\calS_L,\calS_R\},
\end{align*}
and the corresponding restricted predictions
\[
    \widehat{g}^{\mathrm{split}}_{b}(\calS')
    \coloneqq
    \begin{cases}
        1 & \text{if }\widehat{\tau}^{\mathrm{split}}_{b}(\calS') \ge 0,\\
        -1 & \text{if }\widehat{\tau}^{\mathrm{split}}_{b}(\calS') < 0.
    \end{cases}
\]
We score the candidate split using the plug-in estimate of the restricted MSE objective.
Specifically, note that
\[
    \bbE\sqb{\Bigp{\tau_0(X)-g(X)}^2}
    =
    \bbE\sqb{\tau_0(X)^2}+1-2\bbE\sqb{\tau_0(X)g(X)},
\]
so comparing splits amounts to comparing $-\bbE[\tau_0(X)g(X)]$ up to constants.
We therefore use the split-sample score
\begin{align*}
    \widehat{\mathrm{Score}}(\calS_L,\calS_R)
    &\coloneqq
    -\sum_{\calS'\in\{\calS_L,\calS_R\}}
    \frac{\sum_{i\in\calI_b^{\mathrm{split}}}\mathbbm{1}[X_i\in\calS']}{\sum_{i\in\calI_b^{\mathrm{split}}}\mathbbm{1}[X_i\in\calS]}
    \, \widehat{\tau}^{\mathrm{split}}_{b}(\calS')\,\widehat{g}^{\mathrm{split}}_{b}(\calS')\\
    &=
    -\sum_{\calS'\in\{\calS_L,\calS_R\}}
    \frac{\sum_{i\in\calI_b^{\mathrm{split}}}\mathbbm{1}[X_i\in\calS']}{\sum_{i\in\calI_b^{\mathrm{split}}}\mathbbm{1}[X_i\in\calS]}
    \, \Bigl|\widehat{\tau}^{\mathrm{split}}_{b}(\calS')\Bigr|.
\end{align*}
Among candidate splits satisfying the leaf size constraints in the next step, choose the split that minimizes $\widehat{\mathrm{Score}}(\calS_L,\calS_R)$.

\item Impose minimum leaf size constraints using the estimation sample $\calI_b^{\mathrm{est}}$.
We require that each resulting leaf $\calS'$ contains at least $k$ treated and $k$ control observations in $\calI_b^{\mathrm{est}}$,
\begin{align*}
    \sum_{i\in\calI_b^{\mathrm{est}}}\mathbbm{1}[X_i\in\calS', D_i=1] \ge k,
    \qquad
    \sum_{i\in\calI_b^{\mathrm{est}}}\mathbbm{1}[X_i\in\calS', D_i=0] \ge k,
    \qquad \calS'\in\{\calS_L,\calS_R\}.
\end{align*}
Stop splitting when no candidate split satisfies the constraint, or when a preset maximum depth is reached.

\item After the tree structure is fixed, estimate the leafwise CATE using only the estimation sample $\calI_b^{\mathrm{est}}$.
For $x$ that falls into a leaf $\calS_b(x)$ of tree $b$, define
\begin{align*}
    \widehat{\tau}_b(x)
    &\coloneqq
    \frac{\sum_{i\in\calI_b^{\mathrm{est}}}\mathbbm{1}[X_i\in\calS_b(x), D_i=1]Y_i}{\sum_{i\in\calI_b^{\mathrm{est}}}\mathbbm{1}[X_i\in\calS_b(x), D_i=1]}
    -
    \frac{\sum_{i\in\calI_b^{\mathrm{est}}}\mathbbm{1}[X_i\in\calS_b(x), D_i=0]Y_i}{\sum_{i\in\calI_b^{\mathrm{est}}}\mathbbm{1}[X_i\in\calS_b(x), D_i=0]},
\end{align*}
and set
\begin{align*}
    \widehat{g}_b(x)
    \coloneqq
    \begin{cases}
        1 & \text{if }\widehat{\tau}_b(x) \ge 0,\\
        -1 & \text{if }\widehat{\tau}_b(x) < 0.
    \end{cases}
\end{align*}
\end{enumerate}

We use $\calI_b^{\mathrm{split}}$ only to decide the partition, and we use $\calI_b^{\mathrm{est}}$ only to estimate leaf quantities.
This honest design is closely aligned with the goal of policy learning, because the final decision depends on whether $\widehat{\tau}_b(x)$ is above or below $0$.
If the same sample were used for both tasks, the tree can place splits that create leaves whose sign is overly tailored to sample noise, which can amplify sign errors out of sample.

Our splitting rule differs from causal forest in that the partition is chosen to minimize the restricted MSE corresponding to predictors in $\{-1,1\}$.
Concretely, a candidate split is scored by a weighted sum of absolute within-child CATE estimates computed on the split sample.
This directly targets the policy decision boundary, because the optimal restricted predictor within a region is given by the sign of the region's average CATE.

The leaf size constraint is imposed on $\calI_b^{\mathrm{est}}$ because the final leafwise CATE estimator is computed using this sample.
In particular, the sign decision
\begin{align*}
    \widehat{g}_b(x)
    \coloneqq
    \begin{cases}
        1 & \text{if }\widehat{\tau}_b(x) \ge 0,\\
        -1 & \text{if }\widehat{\tau}_b(x) < 0
    \end{cases}
\end{align*}
is sensitive to estimation noise when $\widehat{\tau}_b(x)$ is close to $0$.
Ensuring that both treatment arms are sufficiently represented within each leaf reduces the variance of $\widehat{\tau}_b(x)$ and stabilizes the policy recommendation.

\subsection{Justification}

\paragraph{Partitioning and CATE approximation}
Causal-policy forest is a piecewise constant approximation scheme for the CATE, followed by a sign mapping.
Fix a tree partition $\Omega$ and a leaf $\ell \in \Omega$.
Let $\tau_\ell \coloneqq \bbE[\tau_0(X)\mid X\in\ell]$ denote the leafwise average CATE.
If $\tau_0(\cdot)$ is smooth and the leaves shrink in diameter as $n$ grows, then $\tau_0(x)$ is well approximated by $\tau_\ell$ for $x\in\ell$.
In that regime, the first best policy $x \mapsto \mathbbm{1}[\tau_0(x)\ge 0]$ is well approximated by the leafwise policy $x \mapsto \mathbbm{1}[\tau_\ell \ge 0]$.

Honest estimation stabilizes this approximation.
Because the data used to place splits are separated from the data used to estimate leaf quantities, the resulting leafwise estimates behave like standard within leaf estimators, and the forest aggregation further reduces variance.

\paragraph{Why the splitting rule targets the policy objective}
The equivalence in Section~\ref{sec:equivalence} suggests targeting the squared loss
\begin{align*}
    \bbE\Bigsqb{\bigp{\tau_0(X) - g(X)}^2}
\end{align*}
over predictors $g(\cdot)\in\{-1,1\}$.
For a leafwise constant predictor $g(x)=g_\ell$ for $x\in\ell$, we have
\begin{align*}
    \bbE\Bigsqb{\bigp{\tau_0(X) - g(X)}^2 \mid X\in\ell}
    =
    \bbE\Bigsqb{\tau_0(X)^2 \mid X\in\ell} + 1 - 2 g_\ell \bbE\Bigsqb{\tau_0(X)\mid X\in\ell}.
\end{align*}
Thus the optimal restricted prediction within a leaf is
\begin{align*}
    g_\ell^*
    \in
    \argmax_{g_\ell\in\{-1,1\}} g_\ell \bbE\Bigsqb{\tau_0(X)\mid X\in\ell},
\end{align*}
which yields $g_\ell^* = 1$ when $\tau_\ell \ge 0$ and $g_\ell^*=-1$ when $\tau_\ell < 0$.
Moreover, because $g_\ell^2=1$, comparing partitions amounts to comparing how much the partition increases the magnitude of the leafwise average CATE, through the term $|\bbE[\tau_0(X)\mid X\in\ell]|$.

\paragraph{Comparison to policy tree and Riesz regression} \citet{Zhou2023offlinemultiaction} proposes a policy tree, which trains a policy using a tree-based algorithm and pseudo outcomes.
Unlike that approach, our algorithm does not require constructing a pseudo outcome, which often depends on estimating the propensity score or the Riesz representer. In addition, as noted in Remark~\ref{rem:treeriesz}, our algorithm incorporates Riesz representer estimation through the tree partitioning itself.

\section{Simulation study}
This section reports a simulation study based on the implementation in our accompanying notebook.

\paragraph{Data generating process.}
We generate a synthetic observational dataset with sample size $n=10000$ and covariate dimension $p=10$.
Covariates $X$ are drawn i.i.d.\ from a standard normal distribution.
Treatment assignment is confounded, where the propensity score varies with one coordinate of $X$ through a logistic link, and the CATE varies with two coordinates of $X$, so that treatment effect heterogeneity is present.
Outcomes follow an additive model with a baseline component that depends on $X$, an interaction term $D\tau_0(X)$, and an independent noise term.
In this design, the CATE function is known to the analyst, so we can evaluate learned policies against the oracle benchmark.

\paragraph{Methods compared.}
We compare the following approaches.
First, we report the oracle policy that treats if and only if the true CATE is nonnegative.
Second, we consider a policy tree baseline that uses pseudo outcomes, following the policy learning implementation in the \texttt{causalml} \texttt{PolicyLearner} class, with a depth two decision tree as the policy model, which we denote as Policy tree (DR).
Third, we consider a plug-in approach based on CATE estimation, where we estimate individual treatment effects using the X learner with a gradient boosting regressor base learner, and we convert the estimated CATE into a policy by thresholding at zero, which we denote as X learner.
Finally, we evaluate our causal-policy forest. We implement it through a minimal modification of a causal forest routine, where we basically keep the original forest training procedure and only change outputs in the recursive partition and the final CATE estimation to take values in $\{-1,1\}$.

\paragraph{Evaluation metric.}
Because the DGP is known, we evaluate each learned policy by its population objective up to a constant.
Concretely, for each method we compute the average CATE among units that the learned policy assigns to treatment, which equals the welfare gain relative to assigning treatment $0$ to everyone in this simulation design.
We also report regret, defined as the difference between the oracle policy value and the value attained by the method.

\paragraph{Results.}
Table~\ref{tab:sim_main} summarizes the results from the notebook.
The policy tree baseline improves over the plug-in X learner in this design, but both remain noticeably below the oracle.
In contrast, causal-policy forest attains a policy value close to the oracle benchmark, with a much smaller regret.
This pattern is consistent with the goal of our method, it leverages a tree-based representation to target the policy decision boundary through thresholded leafwise CATE estimates, while keeping the computational advantages of causal forest training.

\begin{table}[t]
\centering
\begin{tabular}{lcc}
\hline
Method & Policy value & Regret \\
\hline
Oracle policy & 0.1833 & 0.0000 \\
Policy tree (DR) & 0.1247 & 0.0586 \\
X learner & 0.0834 & 0.0999 \\
Causal-policy forest & 0.1730 & 0.0103 \\
\hline
\end{tabular}
\caption{Simulation results based on the accompanying notebook.
Policy value is the average CATE among units assigned to treatment by the learned policy, which equals the welfare improvement relative to always assigning treatment $0$ in this DGP.
Regret is computed relative to the oracle policy.}
\label{tab:sim_main}
\end{table}
\section{Conclusion}
This study proposes causal-policy forest, an end-to-end policy learning method that extends causal forest by directly targeting the policy learning objective through a CATE reduction. We show that, under binary treatments, maximizing welfare over a policy class is equivalent to minimizing a squared loss for the CATE when the predictor is restricted to take values in $\{-1,1\}$, and we leverage this equivalence to design a forest algorithm whose splits are chosen to reduce the squared error between the CATE target and a leafwise $\{-1,1\}$ predictor. The resulting procedure learns a policy directly from data while preserving the computational advantages and practical scalability of tree-based methods, since the policy is obtained by simple leafwise aggregation and thresholding rather than by solving a combinatorial welfare maximization problem. Our algorithm also clarifies the close connection between CATE estimation and policy learning, and provides a transparent, modular approach that can incorporate standard forest techniques such as subsampling, random feature selection, and honest estimation, while maintaining a clear interpretation as a piecewise constant approximation of the CATE followed by a sign rule.

\bibliography{arXiv2.bbl}

\bibliographystyle{tmlr}

\onecolumn

\appendix

\end{document}